\documentclass[prl,superscriptaddress,onecolumn]{revtex4}

\usepackage{dcolumn}% Align table columns on decimal point
\usepackage{bm}% bold math
\usepackage{amssymb}
\usepackage{amsmath}
\usepackage{amsfonts}
\usepackage[english]{babel}
\usepackage{amsmath}
\usepackage{graphicx}% Include figure files
\usepackage{dcolumn}% Align table columns on decimal point
\usepackage{bm}% bold math
\usepackage{hyperref}% add hypertext capabilities
\usepackage{bigints}
\usepackage{eqnarray,amsmath}
\usepackage{textcomp}
\usepackage{siunitx}

\RequirePackage{units}
\usepackage[usenames, dvipsnames]{color}

\usepackage[]{graphicx}

\usepackage{epstopdf}
\begin{document}

\title{Anisotropic skyrmion bubbles in ultra-thin epitaxial Au$_{0.67}$Pt$_{0.33}$/Co/W films}
\author{Lorenzo Camosi} \email{lorenzo.camosi@gmail.com}
\affiliation{Univ.~Grenoble Alpes, CNRS, Institut N\'eel, F-38000 Grenoble, France}
\affiliation{Current address: ICN2, Univ. Autonomous of Barcelona, Barcelona, 08193, Spain}
\author{Jose Pe\~na Garcia} \email{jose.pena-garcia@neel.cnrs.fr}
\affiliation{Univ.~Grenoble Alpes, CNRS, Institut N\'eel, F-38000 Grenoble, France}
\author{Olivier Fruchart}
\affiliation{Univ.~Grenoble Alpes, CNRS, CEA, SPINTEC, F-38000 Grenoble, France}
\author{Stefania Pizzini}
\affiliation{Univ.~Grenoble Alpes, CNRS, Institut N\'eel, F-38000 Grenoble, France}
\author{Andrea Locatelli}
\affiliation{Elettra - Sincrotrone Trieste S.C.p.A, S.S 14 - km 163.5 in AREA Science Park 34149 Basovizza, Trieste, Italy}
\author{Tevfik Onur Mente\c{s}}
\affiliation{Elettra - Sincrotrone Trieste S.C.p.A, S.S 14 - km 163.5 in AREA Science Park 34149 Basovizza, Trieste, Italy}
\author{Francesca Genuzio}
\affiliation{Elettra - Sincrotrone Trieste S.C.p.A, S.S 14 - km 163.5 in AREA Science Park 34149 Basovizza, Trieste, Italy}
\author{Justin M. Shaw}
\affiliation{Quantum Electromagnetics Division, National Institute of Standards and Technology, Boulder, CO
80305, USA}
\author{Hans T. Nembach}
\affiliation{Quantum Electromagnetics Division, National Institute of Standards and Technology, Boulder, CO
80305, USA}
\affiliation{JILA, University of Colorado, Boulder, CO 80309, USA}
\author{Jan Vogel} \email{jan.vogel@neel.cnrs.fr}
\affiliation{Univ.~Grenoble Alpes, CNRS, Institut N\'eel, F-38000 Grenoble, France}

%%%%%%%%%%%%%%%%%%%%%%%%%%%%%%%%%%%%%%%%%%%%%%%%%%%%%%%%%%%%%%%%%%%%%
%% The document title should be given as usual. Some journals require
%% a running title from the author: this should be supplied as an
%% optional argument to \title.
%%%%%%%%%%%%%%%%%%%%%%%%%%%%%%%%%%%%%%%%%%%%%%%%%%%%%%%%%%%%%%%%%%%%%

%%%%%%%%%%%%%%%%%%%%%%%%%%%%%%%%%%%%%%%%%%%%%%%%%%%%%%%%%%%%%%%%%%%%%
%% Some journals require a list of abbreviations or keywords to be
%% supplied. These should be set up here, and will be printed after
%% the title and author information, if needed.
%%%%%%%%%%%%%%%%%%%%%%%%%%%%%%%%%%%%%%%%%%%%%%%%%%%%%%%%%%%%%%%%%%%%%

%%%%%%%%%%%%%%%%%%%%%%%%%%%%%%%%%%%%%%%%%%%%%%%%%%%%%%%%%%%%%%%%%%%%%
%% The manuscript does not need to include \maketitle, which is
%% executed automatically.
%%%%%%%%%%%%%%%%%%%%%%%%%%%%%%%%%%%%%%%%%%%%%%%%%%%%%%%%%%%%%%%%%%%%%


%\begin{document}
%\onecolumn
\begin{abstract}
We studied the symmetry of magnetic properties and the resulting magnetic textures in ultra-thin epitaxial Au$_{0.67}$Pt$_{0.33}$/Co/W, a model system exhibiting perpendicular magnetic anisotropy and interface Dzyaloshinskii-Moriya interaction (DMI). As a peculiar feature, the C$_\mathrm{2v}$ crystal symmetry induced by the Co/W interface results in an additional uniaxial in-plane magnetic anisotropy in the cobalt layer. Photoemission electron microscopy with magnetic sensitivity reveals the formation of self-organized magnetic stripe domains oriented parallel to the hard in-plane magnetization axis. We attribute this behavior to the lower domain wall energy when oriented along this axis, where both the DMI and the in-plane magnetic anisotropy favor a N\'{e}el domain wall configuration. The anisotropic domain wall energy also leads to the formation of elliptical skyrmion bubbles in a weak out-of-plane magnetic field.
\end{abstract}

\maketitle
%\twocolumn
%\section{Introduction}

Chiral magnetic spin textures are interesting candidates for future applications in data storage and spintronics \cite{Fert2013}. In perpendicularly magnetized ultrathin films, chirality may be induced by the Dzyaloshinskii-Moriya interaction (DMI) \cite{Dzyaloshinskii1957,Moriya1960}. The DMI is a chiral antisymmetric exchange interaction, which may exist in systems with a lack of structural inversion symmetry and a strong spin-orbit coupling. In ultrathin magnetic films, the conditions that can lead to an interfacial DMI are satisfied if the two interfaces are different (inversion symmetry breaking) and is increased if at least one of the interfaces contains a heavy metal such as Pt or W (strong spin-orbit coupling).

An appropriate balance between the DMI, Heisenberg exchange interaction, dipolar interactions and magnetocrystalline anisotropy (MCA) can lead to the stabilization of chiral N\'eel domain walls (DW) and N\'eel-type skyrmions \cite{Thiaville2012,Rohart2013}. Skyrmions are chiral whirling magnetic configurations with a non-trivial topology \cite{Bogdanov2001}, which have been theoretically investigated \cite{Roessler2006} and experimentally detected in bulk systems with B20 symmetry \cite{Muhlbauer2009}, and, in a metastable state, in ultra-thin magnetic films \cite{Romming2013,Moreau2016,Boulle2016}. Until now, most thin film magnetic systems hosting skyrmions and skyrmion bubbles \cite{Bernand2018} at room temperature were prepared by sputter deposition \cite{Moreau2016,Boulle2016,Jiang2015,Woo2016}, leading to polycrystalline systems with isotropic properties within the film plane. These systems display an in-plane circular symmetry, so that domain walls and skyrmions show isotropic properties.

In epitaxial systems, however, the symmetry of the magnetic interactions reflects the crystal symmetry of the magnetic crystal and its interfaces \cite{Bogdanov1989,Chen2015,Chen2015a}. In-plane strain can lead to distorted skyrmion phases \cite{Shibata2015,Hsu2016}. In systems with C$_\mathrm{2v}$ symmetry, uniaxial and/or biaxial in-plane magnetic anisotropies can occur \cite{Fritzsche1995}, whereas the DMI may have different strength and sign along two perpendicular in-plane directions \cite{Hoffmann2017}. We recently confirmed experimentally such anisotropic properties in epitaxial Au/Co/W(110) thin films with C$_\mathrm{2v}$ symmetry, where in addition to an out-of-plane easy magnetization axis, a uniaxial in-plane anisotropy and a strongly anisotropic DMI were observed \cite{Camosi2017}.

In the present work, we investigate the influence of the out-of-plane and in-plane anisotropies on magnetic textures such as domain walls and skyrmion bubbles in an epitaxial AuPt/Co/W film with C$_\mathrm{2v}$ symmetry. The introduction of Pt in the top layer is expected to strongly increase the DMI with respect to the Au/Co/W system studied previously, while reducing the out-of-plane MCA. The system is grown using pulsed laser deposition (PLD) and the crystal symmetry is studied in-situ with reflection high energy electron diffraction (RHEED). The magnetic properties are investigated with a vibrating-sample magnetometer based on a superconducting quantum interference device (VSM-SQUID), Kerr magnetometry, and Brillouin light scattering (BLS). Chiral domain walls and skyrmions are studied theoretically using micromagnetic simulations and are observed experimentally using photo emission electron microscopy combined with X-ray magnetic circular dichroism (XMCD-PEEM). These measurements show that the anisotropic magnetic properties lead to the formation of magnetic stripe domains spontaneously oriented along the in-plane hard magnetization axis, which transform into elliptical skyrmion bubbles upon application of a small out-of-plane magnetic field.

%\section{Results and discussion}

\begin{figure*}[h!]
\includegraphics[scale=0.35]{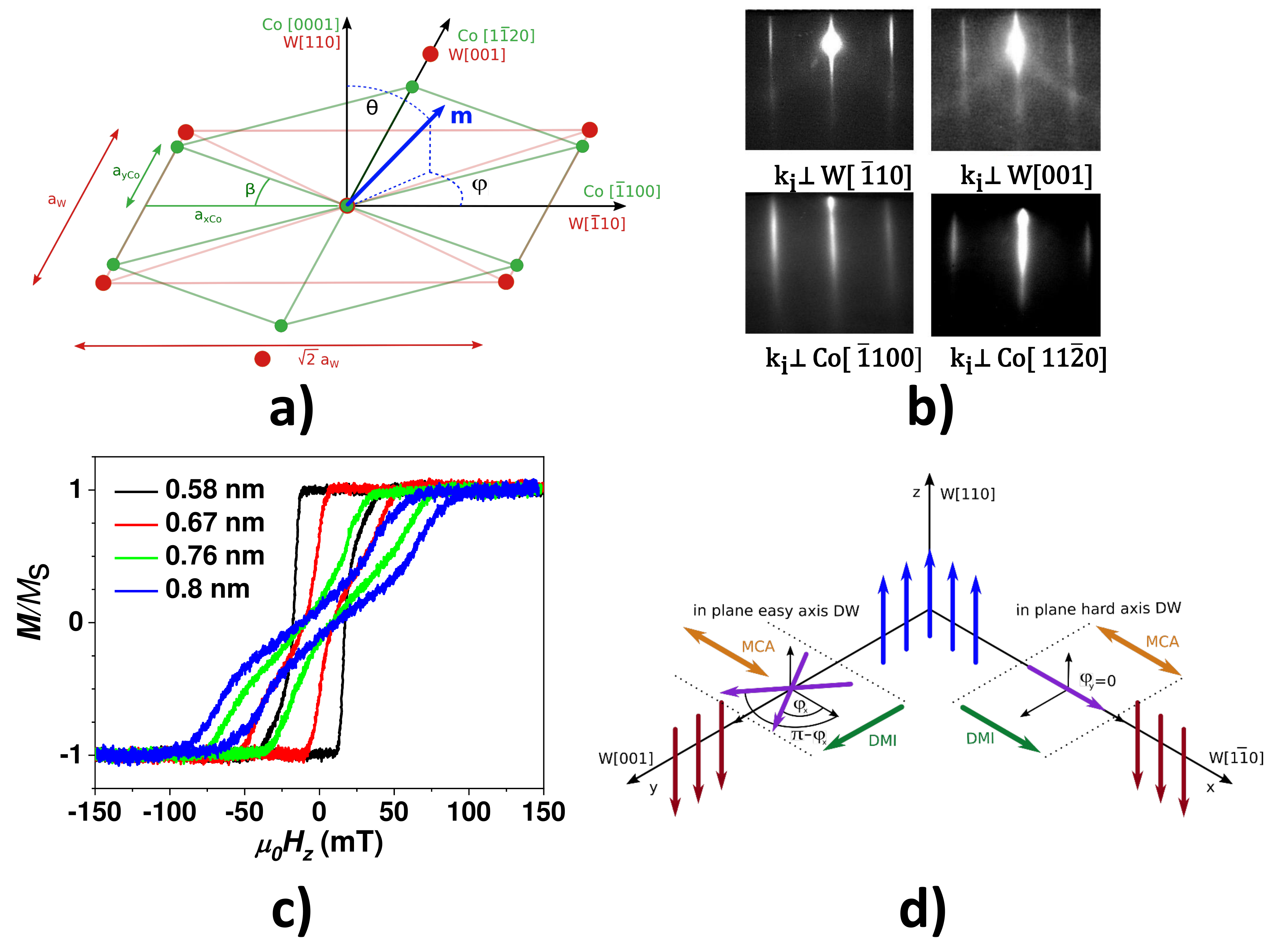}
\caption { a) Overlay of the W(110) and the strained Co(0001) surfaces with the Nishiyama-Wasserman epitaxial relationship~\cite{Nishiyama1934,Wassermann1935}. Red and green dots represent Tungsten and Cobalt atoms. The geometry and notation used to describe the magnetization $\textbf{m}$ in polar coordinates ($\theta$, $\varphi$) in the W bcc(110) crystal frame are also illustrated. b) RHEED diffraction patterns when the incoming electron beam, $\mathrm{k_i}$ is perpendicular to the main in-plane crystallographic axes of W(110) and Co (0001). c) Hysteresis loops obtained by focused polar Kerr measurements for different Co thicknesses. d) Sketch of the possible configurations for domain walls oriented along the in-plane hard axis and the in-plane easy axis. Red and blue arrows : magnetization direction. Orange arrows : effective in-plane anisotropy field direction. Green arrows : effective DMI field }
\label{Figure1}
\end{figure*}

The Au$_{0.67}$Pt$_{0.33}$/Co/W stack was grown in a PLD system under ultra-high vacuum conditions ($P\approx 10^{-8}$Pa) [See Methods]. A wedge of Co with a thickness ranging from 0.4 to 0.8 nanometers (2 to 4 monolayers) was deposited with the help of a computer-controlled mask moving in front of the sample. The Co(0001) film grows on the W bcc(110) surface following the Nishiyama-Wassermann orientation \cite{Nishiyama1934,Wassermann1935}, \textit{i.e.,} with a unique epitaxial relationship with the Co$[1100]$-direction parallel to W$[\bar{1}10]$ and the Co$[11\bar{2}0]$ parallel to W$[001]$ (Fig.~\ref{Figure1}(a)). These epitaxial relationships are confirmed by RHEED (Fig.~\ref{Figure1}(b)), where it is seen that the Co grows pseudomorphically along W$[\bar{1}10]$ but is relaxed along W$[001]$.
The C$_\mathrm{2v}$ symmetry, resulting from the epitaxial growth on W(110), induces an in-plane magnetic anisotropy energy density ($E_\mathrm{MAE}$) in the system, in addition to the uniaxial out-of-plane anisotropy induced by the AuPt/Co interface. $E_\mathrm{MAE}$ may be expressed in the second-order approximation using spherical coordinates $(\theta, \varphi)$ [see Fig.~\ref{Figure1}(a)]

\begin{table*} [t]
\centering
\begin{tabular}{|l|c|c|}
\hline
System & Au/Co/W \cite{Camosi2017} & Au$_{0.67}$Pt$_{0.33}$/Co/W \\
\hline
$D_\mathrm{x}\cdot t_{\mathrm{Co}} \si{(pJ/m)}$ & 0.29$\pm$ 0.03 & 0.74$\pm$ 0.1 \\
\hline
$D_\mathrm{y}\cdot t_{\mathrm{Co}} \si{(pJ/m)}$ & 0.12$\pm$ 0.02 & 0.79$\pm$ 0.1 \\
\hline
$K_\mathrm{u} \si{(kJ/m^3)} $& 1100$\pm$400 & 970$\pm$100\\
\hline
$K_\mathrm{in} \si{(kJ/m^3)}$ & 136 $\pm$ 6 & 94$\pm$ 9  \\
\hline
\end{tabular}
\caption{BLS measurements for Au/Co(0.65nm)/W (from Ref.~\cite{Camosi2017}) and Au$_{0.67}$Pt$_{0.33}$/Co(0.8nm)/W.}
\label{Table1}
\end{table*}

\begin{equation}
E_{\mathrm{MAE}}(\theta,\varphi )=-K_\mathrm{out}\cos^2{\theta}-K_\mathrm{in}\sin^2{\theta}\cos^2{\varphi}
\label{EMAE}
\end{equation}

\noindent where $K_\mathrm{out}$ is the effective out-of-plane anisotropy coefficient, defined as the difference in energy between the magnetization  oriented along the hard in-plane direction and oriented perpendicular to the film plane. It can be expressed as $K_\mathrm{out}$ = $K_\mathrm{u}$ - $K_\mathrm{d}$ with $K_\mathrm{u}$ being the out-of-plane magneto-crystalline anisotropy constant and $K_\mathrm{d}=\dfrac{1}{2}\mu_0M_\mathrm{s}^2$ the dipolar constant. Finally, $K_\mathrm{in}$ is the in-plane anisotropy coefficient, the difference in energy for the magnetization oriented along the in-plane hard and easy axes. VSM-SQUID hysteresis loops with a magnetic field along the main in-plane crystallographic axes showed a smaller saturation field for a magnetic field applied along W$[\bar{1}10]$ than along W$[001]$. Therefore, this confirms the Co axis parallel to W$[\bar{1}10]$ as the in-plane easy axis \cite{Fritzsche1995}. The spontaneous magnetization $M_\mathrm{s}=1.15\pm0.1$~MA/m was extracted from VSM-SQUID measurements for a Co thickness of 0.8~nm.

%We used focused Kerr magnetometry to study the magnetic properties for different Co thicknesses. The polar Kerr cycles were measured as a function of an out-of plane magnetic field.
Figure~\ref{Figure1} (c) shows the hysteresis loops measured  with focused Kerr magnetometry at different locations along the wedge, corresponding to different Co thicknesses. For $t_\mathrm{Co}$ $\approx$ 0.58~nm a square hysteresis loop is obtained, which indicates that the sample magnetization is fully saturated at remanence. With increasing Co thickness, the decrease in the effective anisotropy $K_\mathrm{out}$ leads to a slanted loop, in agreement with a multi-domain state at zero field. The ratio $M/M_\mathrm{S}$ at zero field decreases as $t_\mathrm{Co}$ increases from $\approx$ 0.67~nm to $\approx$ 0.8~nm where the magnetization is close to the spin-reorientation transition (SRT), and the remanence is close to zero.

Table~\ref{Table1} summarizes the BLS results obtained for Au$_{0.67}$Pt$_{0.33}$/Co(0.8nm)/W, compared with our previous results for Au/Co(0.65nm)/W \cite{Camosi2017}. In the following, we will indicate with x the bcc$[\bar{1}10]$ direction (in-plane easy axis) and with y the bcc$[001]$ direction (in-plane hard axis). Since the DMI is mainly an interfacial effect, $D_\mathrm{x}$ and $D_\mathrm{y}$ have been multiplied by $t_\mathrm{Co}$ to take the difference in Co thickness of the two samples into account. The data show that along both in-plane directions the DMI strongly increases with the addition of Pt, compared to Au/Co/W, while the in-plane anisotropy $\mathrm{K_{in}}$ decreases. Note, the error in the value of $D$ mainly originates from the uncertainty in film thickness and therefore in $M_\mathrm{s}$.

\begin{figure*}[t]
\includegraphics[scale=0.4]{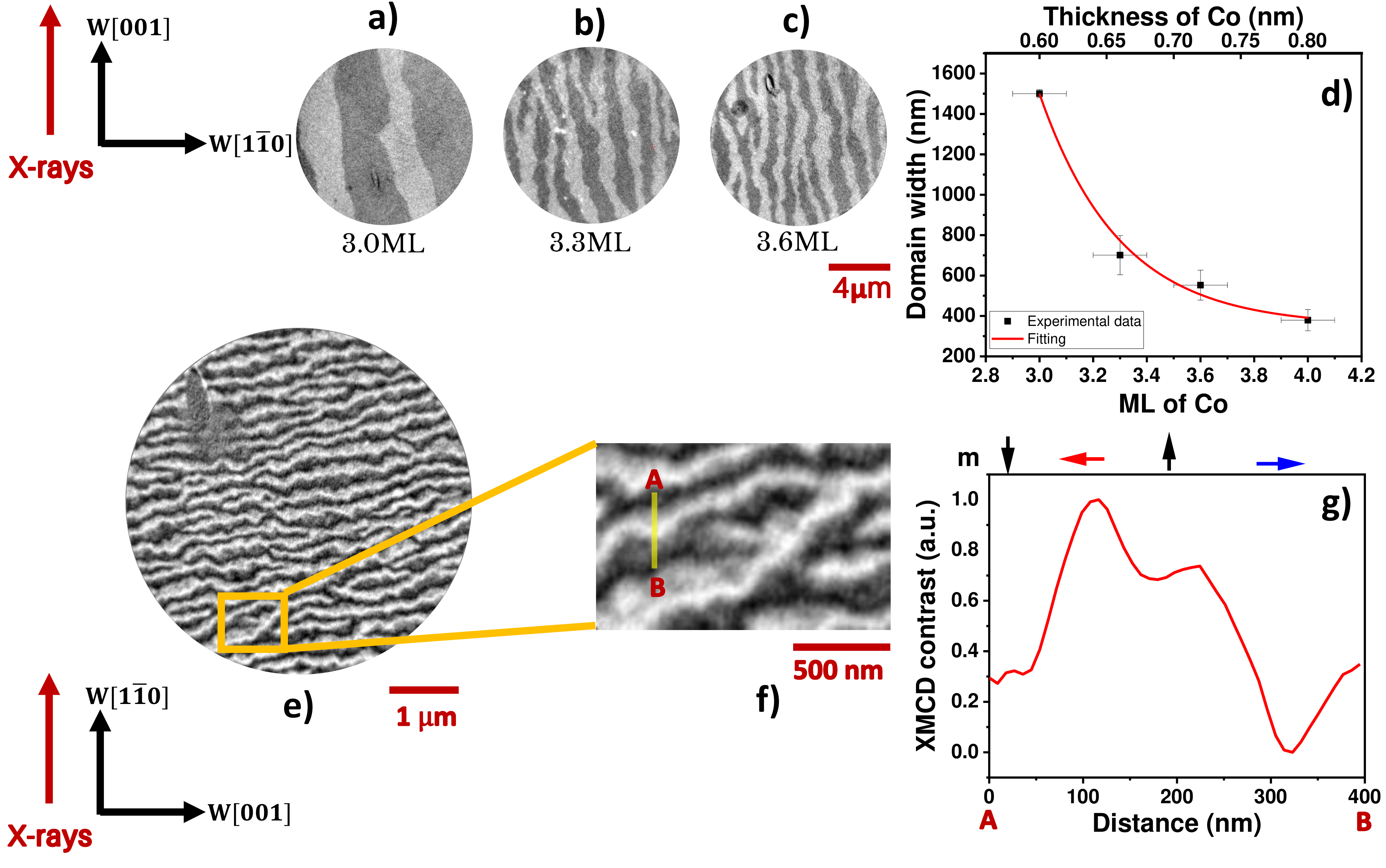}
\caption {a,b,c) 10 $\mu$m diameter XMCD-PEEM images, taken with the projection of the X-ray beam on the sample surface parallel to the W bcc$[001]$ direction, for different thicknesses of Co. The stripe domain width decreases exponentially with the Co thickness as can be seen in d). The thickness of the Co layer was determined from the position in millimeters on the wedge and the known thickness gradient. e) 5 $\mu$m diameter XMCD-PEEM image for 0.8~nm of Co with the X-ray along the W bcc$[1\bar{1}0]$ direction. f) 1.5 $\times$ 1.5 $\mu$m$^2$ zoom of a part of image (e). g) Line scan of the magnetic contrast along the direction perpendicular to the DW direction (along the yellow line in f))}
\label{Figure3}
\end{figure*}

The inclusion of Pt in the top layer strongly increases the DMI strength due to the stronger DMI at the Pt/Co interface \cite{Ajejas2017}, but does not affect the chirality. The measured effective DMI of the trilayer systems is the total contribution from both the top and bottom Co interfaces. It was reported by Ma et al.~\cite{Ma2018} that the DMI at the Co/Au interface is negative, as is also the case for Co/Pt, and stronger than that of Co/W, which is positive. In both Au/Co/W and Pt/Co/W trilayers, the DMI at the two interfaces is thus right-handed, and the positive DMI we measure in our system is in agreement with the literature. The absence of or even slightly opposite anisotropy of the DMI in Au$_\mathrm{0.67}$Pt$_\mathrm{0.33}$/Co/W, with respect to Au/Co/W, can arise partly from the large contribution to the DMI of the Au$_\mathrm{0.67}$Pt$_\mathrm{0.33}$/Co interface, which may be isotropic or even show an anisotropy opposite to the one at the Co/W interface.

In thin film systems with reduced symmetry, the competition between the DMI and the MCA is expected to induce an anisotropic magnetic configuration of the DWs~\cite{Chen2015}. We can expect that for domain walls oriented along the W$[001]$ (in-plane hard axis, y) direction, both the in-plane anisotropy and the DMI are promoting a N\'eel configuration, which is chiral because of the DMI (Fig.~\ref{Figure1} (d)). For domain walls oriented along the W$[\bar{1}10]$ (in-plane easy axis, x) direction, the DMI is promoting a chiral N\'eel configuration while the in-plane anisotropy promotes a non-chiral Bloch configuration (Fig.~\ref{Figure1} (d)), leading to an intermediate angle $\varphi$ (or $\pi$ - $\varphi$), in between a Bloch and a N\'eel wall~\cite{Chen2015}, with $\varphi$ depending on the relative strength of $K_\mathrm{in}$ and $D$. In the phenomenological sketch of Figure~\ref{Figure1} (d) both the DMI and magnetic anisotropy are represented as effective in-plane fields.

\begin{figure*}[t]
\includegraphics[scale=0.35]{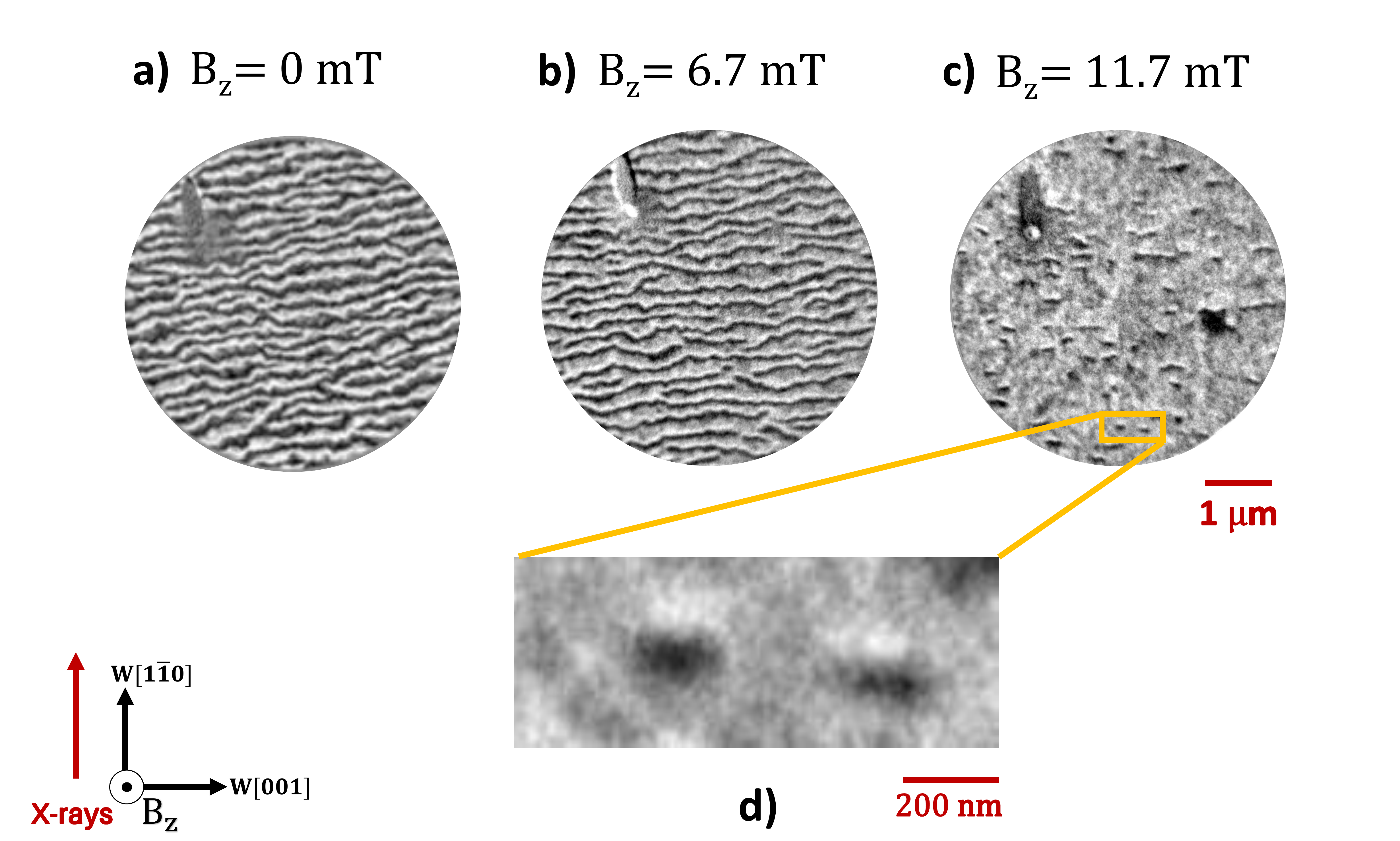}
\caption {XMCD-PEEM images with the X-rays along W $[\bar{1}10]$, at the same position but at different perpendicularly applied magnetic fields. When a magnetic field $\mu_0$H equal to 11.7~mT is applied elliptical skyrmion bubbles are observed.}
\label{Figure4}
\end{figure*}

In order to determine experimentally the domain structure and the domain wall configuration in the Au$_{0.67}$Pt$_{0.33}$/Co/W system, we used XMCD-PEEM. The domain configuration for Co thicknesses between 0.6 and 0.8~nm is shown in Figure~\ref{Figure3} (a,b,c)). As expected from the one-dimensional (1D) model (see Supplemental information), the stripe domains are almost uniquely oriented close to the W$[001]$ direction, as a direct consequence of the lower domain wall energy for domain walls oriented along W$[001]$ than along W$[\bar{1}10]$. The stripe domain width sharply decreases with increasing Co thickness [Fig.~\ref{Figure3}(d)]. This width is determined by the competition between the dipolar energy and the domain wall energy. The experimental points in Figure ~\ref{Figure3}(d) are fitted with an exponential decay, in agreement with theoretical models by Kashuba and Pokrovsky \cite{Kashuba1993}, without considering DMI, and by Meier et al.~\cite{Meier2017} who considered the DMI in a two-dimensional spacing model.

The images in Figure~\ref{Figure3} (a,b,c) do not show any extra magnetic contrast at the domain walls. As predicted by the 1D model, the expected domain wall width is larger than the spatial resolution. However, since the magnetization in the center of the domain walls is expected to be mainly parallel to $[\bar{1}10]$ (Fig.~\ref{Figure1}(d)), it is mainly perpendicular to the incoming x-ray beam and the expected magnetic contrast is in between the contrast of the two domains, and therefore indiscernible. In order to verify this, we rotated the sample by \ang{90} so that the image in Figure~\ref{Figure3}(e) was taken with the X-ray beam oriented along the W $[\bar{1}10]$ direction, for a Co thickness of about 0.8~nm. A clear extra DW contrast is now observed for domain walls oriented along the $[001]$ direction, indicating that the in-plane component of the magnetization is perpendicular to the DW direction. Moreover, the DW contrast alternates between black and white, confirming that the DWs are chiral N\'eel walls. A line scan along the beam direction [Fig.~\ref{Figure3}(f,g)], allows confirming the right-hand chirality of the DWs parallel to the $[001]$ direction, in agreement with the sign of the DMI determined by BLS. From the line scan, a DW width of $\sim$ 52 $\pm$5 nm was determined. This is in good agreement with the value of $\pi\Delta_{y}=59$ nm obtained for the domain wall width along the easy axis from the 1D model. Note that we were not able to determine the DW magnetization for DWs oriented along the in-plane easy axis, since there are almost no domain walls oriented along this direction and they are always very short.

To summarize, in agreement with the 1D model the XMCD-PEEM images reveal a strongly anisotropic domain configuration, with a dominant orientation of stripe domains and domain walls along the $[001]$ direction, the in-plane hard magnetization axis. The images reveal that these domain walls are chiral N\'eel domain walls. The small number and short length of domain walls oriented along the $[\bar{1}10]$ direction did not allow confirming the dominant Bloch configuration expected for such domain walls. However, the strongly anisotropic domain wall orientation is consistent with the difference in magnetic configuration and in domain wall energy of domain walls oriented along the W$[001]$ and W$[\bar{1}10]$ directions. Note that such a strong anisotropy of the domain wall orientation was not observed in the Fe/Ni/W(110) system of Ref.~\cite{Chen2015}, where the value of the DMI was an order of magnitude smaller than in our case.

%In this section, we discuss the effect of the anisotropic energetic environment on the stabilization and shape of magnetic skyrmions and skyrmion bubbles. The difference between a skyrmion and skyrmion bubble is extensively discussed in \cite{Bernand2018}. A skyrmion is stabilized by DMI and exchange interactions, and its size is of the order of a few nanometers \cite{Heinze2011}. On the other hand, a skyrmion bubble is stabilized mainly by dipolar interactions, and its size ranges from tens of nm to a few microns.

\begin{figure*}[h!]
\includegraphics[scale=0.43]{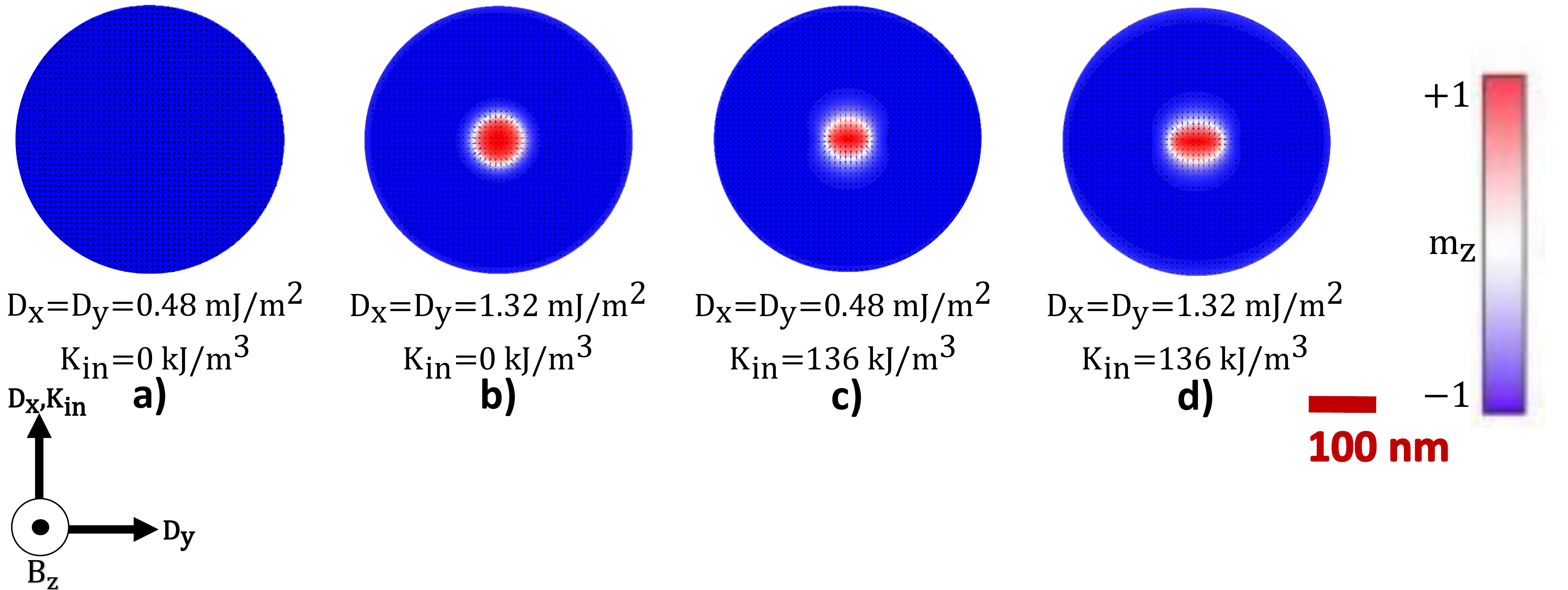}
\caption {Micromagnetic simulations with $K_\mathrm{out}=0.2$~\si{M.J/m^2}, $A=16$~pJ/m, $M_\mathrm{s}=1.15$~MA/m in a 400~nm diameter circular nanodot a) $D_\mathrm{x}=D_\mathrm{y}=0.48$~\si{mJ/m^2}, $K_\mathrm{in}=0$~\si{kJ/m^3} b) $D_\mathrm{x}=D_\mathrm{y}=1.32$~\si{mJ/m^2}, $K_\mathrm{in}=0$~\si{kJ/m^3}, c) $D_\mathrm{x}=D_\mathrm{y}=0.48$~\si{mJ/m^2}, $K_\mathrm{in}=136$~\si{kJ/m^3}, d) $D_\mathrm{x}=D_\mathrm{y}=1.32$~\si{mJ/m^2}, $K_\mathrm{in}=136$~\si{kJ/m^3}}.
\label{Figure5}
\end{figure*}

We discuss now the effect of the anisotropic energetic environment on the stabilization and shape of magnetic skyrmions and skyrmion bubbles. The difference between a skyrmion and skyrmion bubble is extensively discussed in \cite{Bernand2018}. A skyrmion is stabilized by DMI and exchange interactions, and its size is of the order of a few nanometers \cite{Heinze2011}. On the other hand, a skyrmion bubble is stabilized mainly by dipolar interactions, and its size ranges from tens of nm to a few microns. Skyrmion bubbles can be stabilized in thin films starting from a stripe domain phase and applying a perpendicular magnetic field \cite{Jiang2015,Juge2018} or by the confinement in nanodots with the proper lateral size \cite{Rohart2013,Boulle2016}. We investigated the magnetic field-induced skyrmion bubbles in the region close to the SRT ($\sim$ 0.8~nm of Co), where $K_\mathrm{out}$ is small and the presence of metastable skyrmions is favored. Figure~\ref{Figure4} shows the XMCD-PEEM images taken with the x-ray beam parallel to the W $[\bar{1}10]$ direction, under application of different out-of plane magnetic fields. Upon increasing the magnetic field strength, the width of the domains with their magnetization anti-parallel to the field decreases in order to decrease the Zeeman energy [Fig.~\ref{Figure4}(b)]. For strong enough applied fields ($\mu_0H=11.7$~mT), elliptical skyrmion bubbles are observed [Fig.~\ref{Figure4}(c)]. Even larger fields did not allow to stabilize spherical skyrmion bubbles, annihilating the existing bubbles and domains instead. Line scans of the magnetic contrast along the major and minor axes of the skyrmion bubbles give a $R_\mathrm{y}\cong66$~nm and $R_\mathrm{x} \cong23$~nm, where $R_\mathrm{x}$ and $R_\mathrm{y}$ denote the skyrmion radius along the W bcc$[\bar{1}10]$ and $[001]$ directions, respectively.

\begin{figure*}[h!]
\includegraphics[scale=0.32]{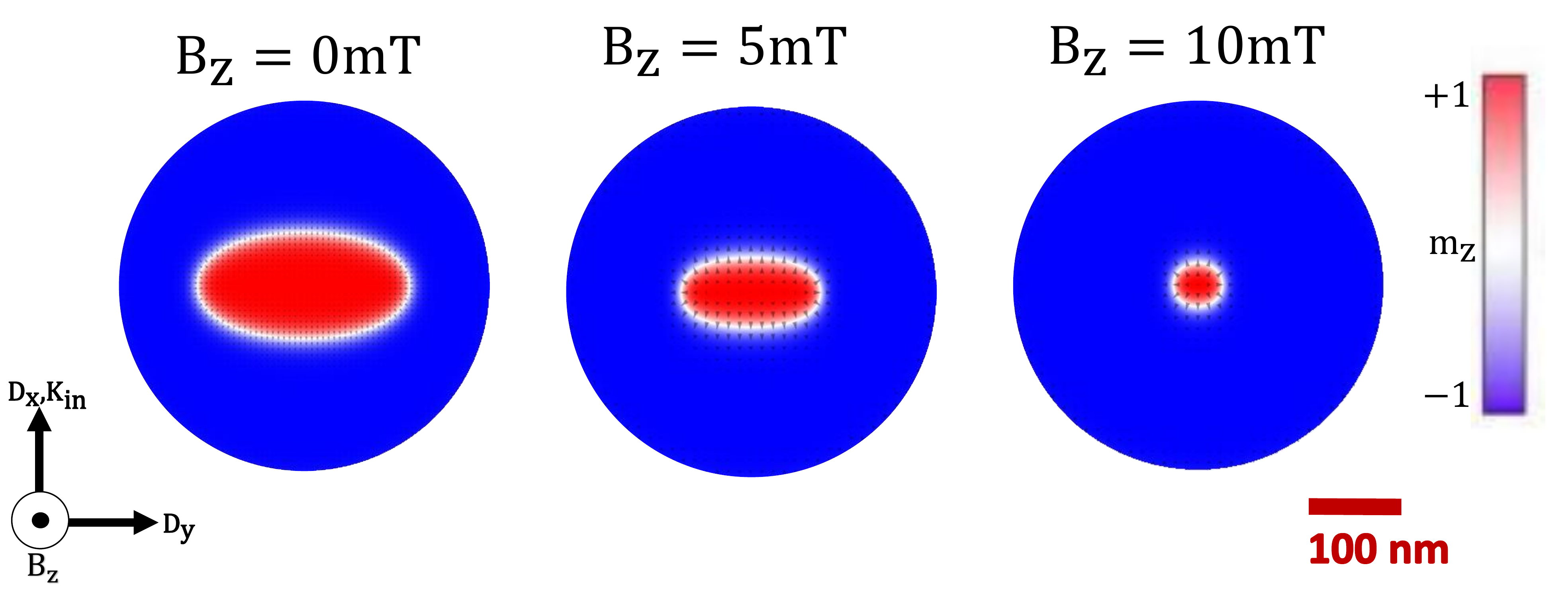}
\caption {Micromagnetic simulations with experimental values of the magnetic parameters reported in Table \ref{Table1} for AuPt/Co/W, for applied perpendicular magnetic fields of $B_\mathrm{z}=0$ mT, 5 mT and 10 mT}
\label{Figure6}
\end{figure*}

In order to understand the role of the in-plane anisotropy on the skyrmion shape, we performed micromagnetic simulations using the code MuMax3 \cite{Vansteenkiste2014}. In the absence of in-plane magnetic anisotropy ($K_\mathrm{in}$ = 0) and for a small DMI value, a skyrmion bubble cannot be stabilized and therefore a uniform ferromagnetic state is obtained [Fig.~\ref{Figure5}(a)]. Using a larger DMI value, a skyrmion bubble can be stabilized [Fig.~\ref{Figure5}(b)]~\cite{Rohart2013}. By adding a non-zero in-plane anisotropy that brings the system closer to an easy-plane anisotropy (\textit{i.e.,} $K_\mathrm{out}=K_\mathrm{in}$), the energy density of domain walls oriented along the hard in-plane axis is reduced, allowing the stabilization of a skyrmion with a small ellipticity for a small DMI value [Fig.~\ref{Figure5}(c)]. The ellipticity increases upon increasing the DMI [Fig.~\ref{Figure5}(d)].

This elliptical shape of the skyrmion bubbles can be explained by the anisotropy of the domain wall energy as discussed before for the domain wall orientation. For a given surface area, the total domain wall energy depends on the shape of the skyrmion bubble, while the surface dipolar energy and the Zeeman energy are more or less constant. The total DW energy, \textit{i.e.,} the DW energy density multiplied by its length, can be minimized by increasing the DW length along the in-plane hard axis and decreasing its DW length along the in-plane easy axis, leading to an elliptical skyrmion shape.

Figure \ref{Figure6} shows the stabilized elliptical skyrmion in a 400nm diameter dot with the experimental values of the magnetic parameters in Table \ref{Table1} for AuPt/Co/W. The best agreement with the experiment is found for an applied magnetic field $B_\mathrm{z}=5$ mT, giving $R_\mathrm{y}\cong65$~nm and $R_\mathrm{x} \cong36$~nm. The magnetic field is smaller than in the experiment ($B_\mathrm{z}=11.7$ mT), but we consider that the agreement is good since the skyrmion size is very sensitive to the magnetic parameters. Moreover, the experiment is performed in a continuous film, while the simulations are for a circular dot, modifying the dipolar effects.

%\section{Conclusions}

In conclusion, we studied the relationship of the symmetry of the crystal and its magnetic properties to the shape of domains and skyrmion bubbles in an epitaxial perpendicularly magnetized AuPt(111)/Co(0001)/W(110) system. We showed that the C$_\mathrm{2v}$ crystal symmetry results in a biaxial magnetic anisotropy and a weakly anisotropic DMI. The change of the capping layer compared to our previous study on Au/Co/W allowed the strength of the magnetic interactions to be tuned while preserving their symmetry. BLS measurements confirmed that the presence of Pt strongly increases the DMI with respect to Au/Co/W.

The interplay between the in-plane anisotropy and the DMI results in a strong dependence of the domain wall energy on the orientation with respect to the crystallographic axes. This anisotropic behavior leads to a spontaneous orientation of magnetic stripe domains along the hard in-plane axis, as revealed by XMCD-PEEM imaging. The domain walls oriented along this axis are right-handed chiral N\'eel domain walls, favored both by the in-plane anisotropy and the DMI. No domain wall contrast occurs for domain walls oriented along the in-plane easy axis, where the competition between the DMI (favoring chiral N\'eel walls) and the in-plane anisotropy (favoring non-chiral Bloch walls) should lead to a mixed Bloch-N\'eel DW configuration.
Our experimental measurements showed that the application of a weak perpendicular magnetic field leads to the transformation of the magnetic stripe domains into skyrmion bubbles with an elliptical shape and a size of about 100~nm. This elliptical shape is due to the anisotropic domain wall energy, in good agreement with micromagnetic simulations. Skyrmion bubbles can be expected to also show strongly anisotropic dynamic properties when driven by spin-polarized currents~\cite{Xia2020}.

\section{Methods}
\subsection{Sample preparation}

The Au$_{0.67}$Pt$_{0.33}$/Co/W stack was grown in a PLD system under Ultra High Vacuum conditions ($P\approx 10^{-8}$Pa). A 10 Hz-pulsed Nd-YAG laser with pulse length $\approx$~10~ns and doubled frequency ($\lambda=532$~nm) was used. The deposition chamber was equipped with a 10~kV RHEED. The substrates were commercial Al$_2$O$_3$($11\bar{2}0)$ single crystals. Prior to the growth, the deposition rate was calibrated using a quartz crystal microbalance.

The growth process is similar to the one described in our previous work \cite{Camosi2017}. First, 0.8~nm of bcc-Mo(110) was deposited at room temperature, followed by the deposition at \si{370 K} of 10~nm bcc-W(110). After annealing at 1070~K for 1~hour, the buffer layer showed high crystal quality as deduced from the in-situ RHEED 1$\times$1 diffraction pattern and the ex-situ $\theta/2\theta$ X-ray diffraction patterns, where Kiessig fringes with many orders were observed (see Supporting Information).
A wedge of Co with a thickness ranging from 0.4 to 0.8 nanometers (2 to 4 monolayers (ML)) was deposited with the help of a computer-controlled mask moving in front of the sample. The sample was progressively warmed from room temperature to 350~K while the Co thickness increased \cite{Camosi2017}.  Finally, a solid solution of Au$_{0.67}$-Pt$_{0.33}$ was deposited at room temperature by alternating seven times the deposition of 0.66~ML of Au and of 0.33~ML of Pt, up to a total thickness of 7~MLs, with the aim to promote the mixture of the two elements at the atomic level and to avoid the formation of misfit dislocations \cite{Fruchart2011}. The deposition is performed at a constant rate of 1~ML/300s.
%\subsection{VSM-SQUID and Kerr magnetometry}

\subsection{BLS measurements}

We performed  BLS spectroscopy in the Damon-Eshbach configuration \cite{Damon1961}, for a Co thickness of 0.8~nm. The magnetization is consecutively saturated along the hard and easy in-plane magnetization axes of the film by an external magnetic field. The spin waves (SW) propagating along the direction perpendicular to this field are probed by a laser with a well-defined wave vector $k_\mathrm{SW}$. Due to the DMI, the SWs involved in the scattering phenomena propagating along opposite directions have different energies. By measuring the spin-wave frequency for positive and negative field polarity, we obtain the frequency-shift $\Delta f=\dfrac{f(+H)-f(-H)}{2}=\dfrac{g\mu_\mathrm{B}}{h}\dfrac{2D}{M_s}{k_\mathrm{SW}}$ with $g$ the Land\'e factor, $\mathrm{\mu_B}$ the Bohr magneton, $h$ Planck's constant, $D$ the micromagnetic DMI constant in \si{J/m^2}, and $M_\mathrm{s}$ the spontaneous magnetization in \si{A/m} \cite{Zakeri2010,Di2015,Nembach2015,Stashkevich2015}. In this experiment, $k_\mathrm{SW}=16.7~\mathrm{\mu m^{-1}}$ and $g=2.17$ \cite{Nembach2015}.

\subsection{XMCD-PEEM}
XMCD-PEEM measurements were performed at the Nanospectroscopy beamline of the Elettra synchrotron (Trieste). The local magnetic contrast in this technique is proportional to the projection of the magnetization on the X-ray beam direction. In our experiment, the X-ray beam is oriented with a grazing incidence of \ang{16} with respect to the sample surface, implying that we are about 3 times more sensitive to the in-plane component of the magnetization than to the out-of-plane component. Together with the high lateral spatial resolution (down to 25 nm) and the high sensitivity, this technique is particularly suited to study the domain wall configuration in our thin film system.

The images were recorded at room temperature at zero magnetic field, with the incoming x-ray beam parallel to the main in-plane crystallographic directions of W(110). Dark and bright grey contrast corresponds to the magnetization pointing up and down, respectively, perpendicular to the film plane.

\subsection{Micromagnetic details}
Micromagnetic simulations were performed with the Micromagnetic code MuMax3\cite{Vansteenkiste2014}. In Figure \ref{Figure5}, we considered an isotropic DMI taking the highest values for Au/Co/W and for AuPt/Co/W in Table \ref{Table1}, divided by the Co thickness of the simulation layer ($t_\mathrm{FM}=0.6$~nm). We used $A=16~\si{pJ/m}$, $K_\mathrm{out}=0.2~\si{MJ/m^3}$, $M_\mathrm{S}=1.15~\si{MA/m}$, and in-plane anisotropy $K_\mathrm{in}=136~\si{kJ/m^3}$, which are the parameters found for Au/Co/W\cite{Camosi2017}. A 60~nm diameter circular bubble domain with zero domain wall width was set in the center of the 400~nm dot and then relaxed under no external magnetic field. An in-plane cellsize of 1 \si{nm} was used.

Figure \ref{Figure6} was obtained considering the experimental magnetic parameters reported in Table \ref{Table1} for AuPt/Co/W, with an exchange stiffness of 16 \si{pJ/m}. The system was minimized under the application of an external perpendicular magnetic field.

\section{Acknowledgements}
LC and JPG contributed equally to this work. JPG acknowledges a PhD grant from the European Commission Horizon 2020 Marie Skłodowska-Curie Actions (Grant no. 754303, GreQue) and the Laboratoire d'excellence LANEF (Grant No. ANR-10-LABX-51-01). LC, JPG, SP and JV acknowledge partial financial support of the Agence Nationale de la Recherche, project ANR-17-CE24-0025 (TOPSKY) and from the DARPA TEE program through Grant No. MIPR HR001183155. JMS and HTN acknowledge partial financial support from the DARPA TEE program through Grant No. MIPR R18-687-0004.

%\bibliographystyle{unsrt}


\bibliography{References}

\begin{thebibliography}{38}
\expandafter\ifx\csname natexlab\endcsname\relax\def\natexlab#1{#1}\fi
\expandafter\ifx\csname bibnamefont\endcsname\relax
  \def\bibnamefont#1{#1}\fi
\expandafter\ifx\csname bibfnamefont\endcsname\relax
  \def\bibfnamefont#1{#1}\fi
\expandafter\ifx\csname citenamefont\endcsname\relax
  \def\citenamefont#1{#1}\fi
\expandafter\ifx\csname url\endcsname\relax
  \def\url#1{\texttt{#1}}\fi
\expandafter\ifx\csname urlprefix\endcsname\relax\def\urlprefix{URL }\fi
\providecommand{\bibinfo}[2]{#2}
\providecommand{\eprint}[2][]{\url{#2}}

\bibitem[{\citenamefont{Fert et~al.}(2013)\citenamefont{Fert, Cros, and
  Sampaio}}]{Fert2013}
\bibinfo{author}{\bibfnamefont{A.}~\bibnamefont{Fert}},
  \bibinfo{author}{\bibfnamefont{V.}~\bibnamefont{Cros}}, \bibnamefont{and}
  \bibinfo{author}{\bibfnamefont{J.}~\bibnamefont{Sampaio}},
  \bibinfo{journal}{Nature Nanotech.} \textbf{\bibinfo{volume}{8}},
  \bibinfo{pages}{152} (\bibinfo{year}{2013}).

\bibitem[{\citenamefont{Dzialoshinskii}(1957)}]{Dzyaloshinskii1957}
\bibinfo{author}{\bibfnamefont{I.}~\bibnamefont{Dzialoshinskii}},
  \bibinfo{journal}{Soviet Physics Jetp-Ussr} \textbf{\bibinfo{volume}{5}},
  \bibinfo{pages}{1259} (\bibinfo{year}{1957}).

\bibitem[{\citenamefont{Moriya}(1960)}]{Moriya1960}
\bibinfo{author}{\bibfnamefont{T.}~\bibnamefont{Moriya}},
  \bibinfo{journal}{Phys. Rev.} \textbf{\bibinfo{volume}{120}},
  \bibinfo{pages}{91} (\bibinfo{year}{1960}).

\bibitem[{\citenamefont{Thiaville et~al.}(2012)\citenamefont{Thiaville, Rohart,
  Ju{\'e}, Cros, and Fert}}]{Thiaville2012}
\bibinfo{author}{\bibfnamefont{A.}~\bibnamefont{Thiaville}},
  \bibinfo{author}{\bibfnamefont{S.}~\bibnamefont{Rohart}},
  \bibinfo{author}{\bibfnamefont{{\'E}.}~\bibnamefont{Ju{\'e}}},
  \bibinfo{author}{\bibfnamefont{V.}~\bibnamefont{Cros}}, \bibnamefont{and}
  \bibinfo{author}{\bibfnamefont{A.}~\bibnamefont{Fert}}, \bibinfo{journal}{EPL
  (Europhysics Letters)} \textbf{\bibinfo{volume}{100}}, \bibinfo{pages}{57002}
  (\bibinfo{year}{2012}).

\bibitem[{\citenamefont{Rohart and Thiaville}(2013)}]{Rohart2013}
\bibinfo{author}{\bibfnamefont{S.}~\bibnamefont{Rohart}} \bibnamefont{and}
  \bibinfo{author}{\bibfnamefont{A.}~\bibnamefont{Thiaville}},
  \bibinfo{journal}{Phys. Rev. B} \textbf{\bibinfo{volume}{88}},
  \bibinfo{pages}{184422} (\bibinfo{year}{2013}).

\bibitem[{\citenamefont{Bogdanov and R{\"o}{\ss}ler}(2001)}]{Bogdanov2001}
\bibinfo{author}{\bibfnamefont{A.}~\bibnamefont{Bogdanov}} \bibnamefont{and}
  \bibinfo{author}{\bibfnamefont{U.}~\bibnamefont{R{\"o}{\ss}ler}},
  \bibinfo{journal}{Phys. Rev. Lett.} \textbf{\bibinfo{volume}{87}},
  \bibinfo{pages}{037203} (\bibinfo{year}{2001}).

\bibitem[{\citenamefont{Roessler et~al.}({2006})\citenamefont{Roessler,
  Bogdanov, and Pfleiderer}}]{Roessler2006}
\bibinfo{author}{\bibfnamefont{U.~K.} \bibnamefont{Roessler}},
  \bibinfo{author}{\bibfnamefont{A.~N.} \bibnamefont{Bogdanov}},
  \bibnamefont{and}
  \bibinfo{author}{\bibfnamefont{C.}~\bibnamefont{Pfleiderer}},
  \bibinfo{journal}{Nature} \textbf{\bibinfo{volume}{{442}}},
  \bibinfo{pages}{797} (\bibinfo{year}{{2006}}).

\bibitem[{\citenamefont{M{\"u}hlbauer et~al.}(2009)\citenamefont{M{\"u}hlbauer,
  Binz, Jonietz, Pfleiderer, Rosch, Neubauer, Georgii, and
  B{\"o}ni}}]{Muhlbauer2009}
\bibinfo{author}{\bibfnamefont{S.}~\bibnamefont{M{\"u}hlbauer}},
  \bibinfo{author}{\bibfnamefont{B.}~\bibnamefont{Binz}},
  \bibinfo{author}{\bibfnamefont{F.}~\bibnamefont{Jonietz}},
  \bibinfo{author}{\bibfnamefont{C.}~\bibnamefont{Pfleiderer}},
  \bibinfo{author}{\bibfnamefont{A.}~\bibnamefont{Rosch}},
  \bibinfo{author}{\bibfnamefont{A.}~\bibnamefont{Neubauer}},
  \bibinfo{author}{\bibfnamefont{R.}~\bibnamefont{Georgii}}, \bibnamefont{and}
  \bibinfo{author}{\bibfnamefont{P.}~\bibnamefont{B{\"o}ni}},
  \bibinfo{journal}{Science} \textbf{\bibinfo{volume}{323}},
  \bibinfo{pages}{915} (\bibinfo{year}{2009}).

\bibitem[{\citenamefont{Romming et~al.}(2013)\citenamefont{Romming, Hanneken,
  Menzel, Bickel, Wolter, von Bergmann, Kubetzka, and
  Wiesendanger}}]{Romming2013}
\bibinfo{author}{\bibfnamefont{N.}~\bibnamefont{Romming}},
  \bibinfo{author}{\bibfnamefont{C.}~\bibnamefont{Hanneken}},
  \bibinfo{author}{\bibfnamefont{M.}~\bibnamefont{Menzel}},
  \bibinfo{author}{\bibfnamefont{J.~E.} \bibnamefont{Bickel}},
  \bibinfo{author}{\bibfnamefont{B.}~\bibnamefont{Wolter}},
  \bibinfo{author}{\bibfnamefont{K.}~\bibnamefont{von Bergmann}},
  \bibinfo{author}{\bibfnamefont{A.}~\bibnamefont{Kubetzka}}, \bibnamefont{and}
  \bibinfo{author}{\bibfnamefont{R.}~\bibnamefont{Wiesendanger}},
  \bibinfo{journal}{Science} \textbf{\bibinfo{volume}{341}},
  \bibinfo{pages}{636} (\bibinfo{year}{2013}).

\bibitem[{\citenamefont{Moreau-Luchaire
  et~al.}(2016)\citenamefont{Moreau-Luchaire, Moutafis, Reyren, Sampaio, Vaz,
  Van~Horne, Bouzehouane, Garcia, Deranlot, Warnicke et~al.}}]{Moreau2016}
\bibinfo{author}{\bibfnamefont{C.}~\bibnamefont{Moreau-Luchaire}},
  \bibinfo{author}{\bibfnamefont{C.}~\bibnamefont{Moutafis}},
  \bibinfo{author}{\bibfnamefont{N.}~\bibnamefont{Reyren}},
  \bibinfo{author}{\bibfnamefont{J.}~\bibnamefont{Sampaio}},
  \bibinfo{author}{\bibfnamefont{C.}~\bibnamefont{Vaz}},
  \bibinfo{author}{\bibfnamefont{N.}~\bibnamefont{Van~Horne}},
  \bibinfo{author}{\bibfnamefont{K.}~\bibnamefont{Bouzehouane}},
  \bibinfo{author}{\bibfnamefont{K.}~\bibnamefont{Garcia}},
  \bibinfo{author}{\bibfnamefont{C.}~\bibnamefont{Deranlot}},
  \bibinfo{author}{\bibfnamefont{P.}~\bibnamefont{Warnicke}},
  \bibnamefont{et~al.}, \bibinfo{journal}{Nature Nanotech.}
  \textbf{\bibinfo{volume}{11}}, \bibinfo{pages}{444} (\bibinfo{year}{2016}).

\bibitem[{\citenamefont{Boulle et~al.}(2016)\citenamefont{Boulle, Vogel, Yang,
  Pizzini, de~Souza~Chaves, Locatelli, Mente{\c{s}}, Sala, Buda-Prejbeanu,
  Klein et~al.}}]{Boulle2016}
\bibinfo{author}{\bibfnamefont{O.}~\bibnamefont{Boulle}},
  \bibinfo{author}{\bibfnamefont{J.}~\bibnamefont{Vogel}},
  \bibinfo{author}{\bibfnamefont{H.}~\bibnamefont{Yang}},
  \bibinfo{author}{\bibfnamefont{S.}~\bibnamefont{Pizzini}},
  \bibinfo{author}{\bibfnamefont{D.}~\bibnamefont{de~Souza~Chaves}},
  \bibinfo{author}{\bibfnamefont{A.}~\bibnamefont{Locatelli}},
  \bibinfo{author}{\bibfnamefont{T.}~\bibnamefont{Mente{\c{s}}}},
  \bibinfo{author}{\bibfnamefont{A.}~\bibnamefont{Sala}},
  \bibinfo{author}{\bibfnamefont{L.}~\bibnamefont{Buda-Prejbeanu}},
  \bibinfo{author}{\bibfnamefont{O.}~\bibnamefont{Klein}},
  \bibnamefont{et~al.}, \bibinfo{journal}{Nature Nanotech.}
  \textbf{\bibinfo{volume}{11}}, \bibinfo{pages}{449} (\bibinfo{year}{2016}).

\bibitem[{\citenamefont{Bernand-Mantel
  et~al.}(2018)\citenamefont{Bernand-Mantel, Camosi, Wartelle, Rougemaille,
  Darques, and Ranno}}]{Bernand2018}
\bibinfo{author}{\bibfnamefont{A.}~\bibnamefont{Bernand-Mantel}},
  \bibinfo{author}{\bibfnamefont{L.}~\bibnamefont{Camosi}},
  \bibinfo{author}{\bibfnamefont{A.}~\bibnamefont{Wartelle}},
  \bibinfo{author}{\bibfnamefont{N.}~\bibnamefont{Rougemaille}},
  \bibinfo{author}{\bibfnamefont{M.}~\bibnamefont{Darques}}, \bibnamefont{and}
  \bibinfo{author}{\bibfnamefont{L.}~\bibnamefont{Ranno}},
  \bibinfo{journal}{SciPost Phys.} \textbf{\bibinfo{volume}{4}},
  \bibinfo{pages}{27} (\bibinfo{year}{2018}).

\bibitem[{\citenamefont{Jiang et~al.}({2015})\citenamefont{Jiang, Upadhyaya,
  Zhang, Yu, Jungfleisch, Fradin, Pearson, Tserkovnyak, Wang, Heinonen
  et~al.}}]{Jiang2015}
\bibinfo{author}{\bibfnamefont{W.}~\bibnamefont{Jiang}},
  \bibinfo{author}{\bibfnamefont{P.}~\bibnamefont{Upadhyaya}},
  \bibinfo{author}{\bibfnamefont{W.}~\bibnamefont{Zhang}},
  \bibinfo{author}{\bibfnamefont{G.}~\bibnamefont{Yu}},
  \bibinfo{author}{\bibfnamefont{M.}~\bibnamefont{Jungfleisch}},
  \bibinfo{author}{\bibfnamefont{F.}~\bibnamefont{Fradin}},
  \bibinfo{author}{\bibfnamefont{J.}~\bibnamefont{Pearson}},
  \bibinfo{author}{\bibfnamefont{Y.}~\bibnamefont{Tserkovnyak}},
  \bibinfo{author}{\bibfnamefont{K.}~\bibnamefont{Wang}},
  \bibinfo{author}{\bibfnamefont{O.}~\bibnamefont{Heinonen}},
  \bibnamefont{et~al.}, \bibinfo{journal}{Science}
  \textbf{\bibinfo{volume}{{349}}}, \bibinfo{pages}{283}
  (\bibinfo{year}{{2015}}).

\bibitem[{\citenamefont{Woo et~al.}({2016})\citenamefont{Woo, Litzius, Krueger,
  Im, Caretta, Richter, Mann, Krone, Reeve, Weigand et~al.}}]{Woo2016}
\bibinfo{author}{\bibfnamefont{S.}~\bibnamefont{Woo}},
  \bibinfo{author}{\bibfnamefont{K.}~\bibnamefont{Litzius}},
  \bibinfo{author}{\bibfnamefont{B.}~\bibnamefont{Krueger}},
  \bibinfo{author}{\bibfnamefont{M.-Y.} \bibnamefont{Im}},
  \bibinfo{author}{\bibfnamefont{L.}~\bibnamefont{Caretta}},
  \bibinfo{author}{\bibfnamefont{K.}~\bibnamefont{Richter}},
  \bibinfo{author}{\bibfnamefont{M.}~\bibnamefont{Mann}},
  \bibinfo{author}{\bibfnamefont{A.}~\bibnamefont{Krone}},
  \bibinfo{author}{\bibfnamefont{R.~M.} \bibnamefont{Reeve}},
  \bibinfo{author}{\bibfnamefont{M.}~\bibnamefont{Weigand}},
  \bibnamefont{et~al.}, \bibinfo{journal}{Nature Mater.}
  \textbf{\bibinfo{volume}{{15}}}, \bibinfo{pages}{{501}}
  (\bibinfo{year}{{2016}}).

\bibitem[{\citenamefont{Bogdanov and Yablonski}(1989)}]{Bogdanov1989}
\bibinfo{author}{\bibfnamefont{A.}~\bibnamefont{Bogdanov}} \bibnamefont{and}
  \bibinfo{author}{\bibfnamefont{D.}~\bibnamefont{Yablonski}},
  \bibinfo{journal}{Zh. Eksp. Teor. Fiz.} \textbf{\bibinfo{volume}{96}},
  \bibinfo{pages}{253} (\bibinfo{year}{1989}).

\bibitem[{\citenamefont{Chen et~al.}({2015})\citenamefont{Chen, N'Diaye, Kang,
  Kwon, Won, Wu, Qiu, and Schmid}}]{Chen2015}
\bibinfo{author}{\bibfnamefont{G.}~\bibnamefont{Chen}},
  \bibinfo{author}{\bibfnamefont{A.~T.} \bibnamefont{N'Diaye}},
  \bibinfo{author}{\bibfnamefont{S.~P.} \bibnamefont{Kang}},
  \bibinfo{author}{\bibfnamefont{H.~Y.} \bibnamefont{Kwon}},
  \bibinfo{author}{\bibfnamefont{C.}~\bibnamefont{Won}},
  \bibinfo{author}{\bibfnamefont{Y.}~\bibnamefont{Wu}},
  \bibinfo{author}{\bibfnamefont{Z.~Q.} \bibnamefont{Qiu}}, \bibnamefont{and}
  \bibinfo{author}{\bibfnamefont{A.~K.} \bibnamefont{Schmid}},
  \bibinfo{journal}{Nature Commun.} \textbf{\bibinfo{volume}{{6}}}
  (\bibinfo{year}{{2015}}).

\bibitem[{\citenamefont{Chen et~al.}(2015)\citenamefont{Chen, Mascaraque,
  N'Diaye, and Schmid}}]{Chen2015a}
\bibinfo{author}{\bibfnamefont{G.}~\bibnamefont{Chen}},
  \bibinfo{author}{\bibfnamefont{A.}~\bibnamefont{Mascaraque}},
  \bibinfo{author}{\bibfnamefont{A.~T.} \bibnamefont{N'Diaye}},
  \bibnamefont{and} \bibinfo{author}{\bibfnamefont{A.~K.}
  \bibnamefont{Schmid}}, \bibinfo{journal}{Applied Physics Letters}
  \textbf{\bibinfo{volume}{106}}, \bibinfo{pages}{242404}
  (\bibinfo{year}{2015}), \urlprefix\url{https://doi.org/10.1063/1.4922726}.

\bibitem[{\citenamefont{Shibata et~al.}(2015)\citenamefont{Shibata, Iwasaki,
  Kanazawa, Aizawa, Tanigaki, Shirai, Nakajima, Kubota, Kawasaki, Park
  et~al.}}]{Shibata2015}
\bibinfo{author}{\bibfnamefont{K.}~\bibnamefont{Shibata}},
  \bibinfo{author}{\bibfnamefont{J.}~\bibnamefont{Iwasaki}},
  \bibinfo{author}{\bibfnamefont{N.}~\bibnamefont{Kanazawa}},
  \bibinfo{author}{\bibfnamefont{S.}~\bibnamefont{Aizawa}},
  \bibinfo{author}{\bibfnamefont{T.}~\bibnamefont{Tanigaki}},
  \bibinfo{author}{\bibfnamefont{M.}~\bibnamefont{Shirai}},
  \bibinfo{author}{\bibfnamefont{T.}~\bibnamefont{Nakajima}},
  \bibinfo{author}{\bibfnamefont{M.}~\bibnamefont{Kubota}},
  \bibinfo{author}{\bibfnamefont{M.}~\bibnamefont{Kawasaki}},
  \bibinfo{author}{\bibfnamefont{H.~S.} \bibnamefont{Park}},
  \bibnamefont{et~al.}, \bibinfo{journal}{Nature Nanotechn.}
  \textbf{\bibinfo{volume}{10}}, \bibinfo{pages}{589} (\bibinfo{year}{2015}).

\bibitem[{\citenamefont{Hsu et~al.}(2016)\citenamefont{Hsu, Kubetzka, Finco,
  Romming, von Bergmann, and Wiesendanger}}]{Hsu2016}
\bibinfo{author}{\bibfnamefont{P.-J.} \bibnamefont{Hsu}},
  \bibinfo{author}{\bibfnamefont{A.}~\bibnamefont{Kubetzka}},
  \bibinfo{author}{\bibfnamefont{A.}~\bibnamefont{Finco}},
  \bibinfo{author}{\bibfnamefont{N.}~\bibnamefont{Romming}},
  \bibinfo{author}{\bibfnamefont{K.}~\bibnamefont{von Bergmann}},
  \bibnamefont{and}
  \bibinfo{author}{\bibfnamefont{R.}~\bibnamefont{Wiesendanger}},
  \bibinfo{journal}{Nature Nanotechnology} \textbf{\bibinfo{volume}{12}},
  \bibinfo{pages}{123} (\bibinfo{year}{2016}).

\bibitem[{\citenamefont{Fritzsche et~al.}(1995)\citenamefont{Fritzsche,
  Kohlhepp, and Gradmann}}]{Fritzsche1995}
\bibinfo{author}{\bibfnamefont{H.}~\bibnamefont{Fritzsche}},
  \bibinfo{author}{\bibfnamefont{J.}~\bibnamefont{Kohlhepp}}, \bibnamefont{and}
  \bibinfo{author}{\bibfnamefont{U.}~\bibnamefont{Gradmann}},
  \bibinfo{journal}{Phys. Rev. B} \textbf{\bibinfo{volume}{51}},
  \bibinfo{pages}{15933} (\bibinfo{year}{1995}).

\bibitem[{\citenamefont{Hoffmann et~al.}(2017)\citenamefont{Hoffmann,
  Zimmermann, M{\"u}ller, Sch{\"u}rhoff, Kiselev, Melcher, and
  Bl{\"u}gel}}]{Hoffmann2017}
\bibinfo{author}{\bibfnamefont{M.}~\bibnamefont{Hoffmann}},
  \bibinfo{author}{\bibfnamefont{B.}~\bibnamefont{Zimmermann}},
  \bibinfo{author}{\bibfnamefont{G.}~\bibnamefont{M{\"u}ller}},
  \bibinfo{author}{\bibfnamefont{D.}~\bibnamefont{Sch{\"u}rhoff}},
  \bibinfo{author}{\bibfnamefont{N.}~\bibnamefont{Kiselev}},
  \bibinfo{author}{\bibfnamefont{C.}~\bibnamefont{Melcher}}, \bibnamefont{and}
  \bibinfo{author}{\bibfnamefont{S.}~\bibnamefont{Bl{\"u}gel}},
  \bibinfo{journal}{Nature Comm.} \textbf{\bibinfo{volume}{8}},
  \bibinfo{pages}{308} (\bibinfo{year}{2017}).

\bibitem[{\citenamefont{Camosi et~al.}(2017)\citenamefont{Camosi, Rohart,
  Fruchart, Pizzini, Belmeguenai, Roussign{\'e}, Stashkevich, Cherif, Ranno,
  De~Santis et~al.}}]{Camosi2017}
\bibinfo{author}{\bibfnamefont{L.}~\bibnamefont{Camosi}},
  \bibinfo{author}{\bibfnamefont{S.}~\bibnamefont{Rohart}},
  \bibinfo{author}{\bibfnamefont{O.}~\bibnamefont{Fruchart}},
  \bibinfo{author}{\bibfnamefont{S.}~\bibnamefont{Pizzini}},
  \bibinfo{author}{\bibfnamefont{M.}~\bibnamefont{Belmeguenai}},
  \bibinfo{author}{\bibfnamefont{Y.}~\bibnamefont{Roussign{\'e}}},
  \bibinfo{author}{\bibfnamefont{A.}~\bibnamefont{Stashkevich}},
  \bibinfo{author}{\bibfnamefont{S.}~\bibnamefont{Cherif}},
  \bibinfo{author}{\bibfnamefont{L.}~\bibnamefont{Ranno}},
  \bibinfo{author}{\bibfnamefont{M.}~\bibnamefont{De~Santis}},
  \bibnamefont{et~al.}, \bibinfo{journal}{Phys. Rev. B}
  \textbf{\bibinfo{volume}{95}}, \bibinfo{pages}{214422}
  (\bibinfo{year}{2017}).

\bibitem[{\citenamefont{Nishiyama}(1934)}]{Nishiyama1934}
\bibinfo{author}{\bibfnamefont{Z.}~\bibnamefont{Nishiyama}},
  \bibinfo{journal}{Sci. Rep. Tohoku Univ.} \textbf{\bibinfo{volume}{23}},
  \bibinfo{pages}{637} (\bibinfo{year}{1934}).

\bibitem[{\citenamefont{Wassermann}(1935)}]{Wassermann1935}
\bibinfo{author}{\bibfnamefont{G.}~\bibnamefont{Wassermann}},
  \emph{\bibinfo{title}{Ueber den Mechanismus der [alpha]-[gamma]-Umwandlung
  des Eisens}} (\bibinfo{publisher}{Verlag Stahleisen}, \bibinfo{year}{1935}).

\bibitem[{\citenamefont{Ajejas et~al.}(2017)\citenamefont{Ajejas,
  K{\v{r}}i{\v{z}}{\'a}kov{\'a}, de~Souza~Chaves, Vogel, Perna, Guerrero,
  Gudin, Camarero, and Pizzini}}]{Ajejas2017}
\bibinfo{author}{\bibfnamefont{F.}~\bibnamefont{Ajejas}},
  \bibinfo{author}{\bibfnamefont{V.}~\bibnamefont{K{\v{r}}i{\v{z}}{\'a}kov{\'a}}},
  \bibinfo{author}{\bibfnamefont{D.}~\bibnamefont{de~Souza~Chaves}},
  \bibinfo{author}{\bibfnamefont{J.}~\bibnamefont{Vogel}},
  \bibinfo{author}{\bibfnamefont{P.}~\bibnamefont{Perna}},
  \bibinfo{author}{\bibfnamefont{R.}~\bibnamefont{Guerrero}},
  \bibinfo{author}{\bibfnamefont{A.}~\bibnamefont{Gudin}},
  \bibinfo{author}{\bibfnamefont{J.}~\bibnamefont{Camarero}}, \bibnamefont{and}
  \bibinfo{author}{\bibfnamefont{S.}~\bibnamefont{Pizzini}},
  \bibinfo{journal}{Appl. Phys. Lett.} \textbf{\bibinfo{volume}{111}},
  \bibinfo{pages}{202402} (\bibinfo{year}{2017}).

\bibitem[{\citenamefont{Ma et~al.}(2018)\citenamefont{Ma, Yu, Tang, Li, He,
  Shi, Wang, and Li}}]{Ma2018}
\bibinfo{author}{\bibfnamefont{X.}~\bibnamefont{Ma}},
  \bibinfo{author}{\bibfnamefont{G.}~\bibnamefont{Yu}},
  \bibinfo{author}{\bibfnamefont{C.}~\bibnamefont{Tang}},
  \bibinfo{author}{\bibfnamefont{X.}~\bibnamefont{Li}},
  \bibinfo{author}{\bibfnamefont{C.}~\bibnamefont{He}},
  \bibinfo{author}{\bibfnamefont{J.}~\bibnamefont{Shi}},
  \bibinfo{author}{\bibfnamefont{K.~L.} \bibnamefont{Wang}}, \bibnamefont{and}
  \bibinfo{author}{\bibfnamefont{X.}~\bibnamefont{Li}}, \bibinfo{journal}{Phys.
  Rev. Lett.} \textbf{\bibinfo{volume}{120}}, \bibinfo{pages}{157204}
  (\bibinfo{year}{2018}).

\bibitem[{\citenamefont{Kashuba and Pokrovsky}(1993)}]{Kashuba1993}
\bibinfo{author}{\bibfnamefont{A.}~\bibnamefont{Kashuba}} \bibnamefont{and}
  \bibinfo{author}{\bibfnamefont{V.~L.} \bibnamefont{Pokrovsky}},
  \bibinfo{journal}{Phys. Rev. B} \textbf{\bibinfo{volume}{48}},
  \bibinfo{pages}{10335} (\bibinfo{year}{1993}).

\bibitem[{\citenamefont{Meier et~al.}(2017)\citenamefont{Meier, Kronseder, and
  Back}}]{Meier2017}
\bibinfo{author}{\bibfnamefont{T.}~\bibnamefont{Meier}},
  \bibinfo{author}{\bibfnamefont{M.}~\bibnamefont{Kronseder}},
  \bibnamefont{and} \bibinfo{author}{\bibfnamefont{C.}~\bibnamefont{Back}},
  \bibinfo{journal}{Phys. Rev. B} \textbf{\bibinfo{volume}{96}},
  \bibinfo{pages}{144408} (\bibinfo{year}{2017}).

\bibitem[{\citenamefont{Heinze et~al.}({2011})\citenamefont{Heinze, von
  Bergmann, Menzel, Brede, Kubetzka, Wiesendanger, Bihlmayer, and
  Bluegel}}]{Heinze2011}
\bibinfo{author}{\bibfnamefont{S.}~\bibnamefont{Heinze}},
  \bibinfo{author}{\bibfnamefont{K.}~\bibnamefont{von Bergmann}},
  \bibinfo{author}{\bibfnamefont{M.}~\bibnamefont{Menzel}},
  \bibinfo{author}{\bibfnamefont{J.}~\bibnamefont{Brede}},
  \bibinfo{author}{\bibfnamefont{A.}~\bibnamefont{Kubetzka}},
  \bibinfo{author}{\bibfnamefont{R.}~\bibnamefont{Wiesendanger}},
  \bibinfo{author}{\bibfnamefont{G.}~\bibnamefont{Bihlmayer}},
  \bibnamefont{and} \bibinfo{author}{\bibfnamefont{S.}~\bibnamefont{Bluegel}},
  \bibinfo{journal}{{Nature Phys.}} \textbf{\bibinfo{volume}{{7}}},
  \bibinfo{pages}{713} (\bibinfo{year}{{2011}}).

\bibitem[{\citenamefont{Juge et~al.}(2018)\citenamefont{Juge, Je,
  de~Souza~Chaves, Pizzini, Buda-Prejbeanu, Aballe, Foerster, Locatelli,
  Mente{\c{s}}, Sala et~al.}}]{Juge2018}
\bibinfo{author}{\bibfnamefont{R.}~\bibnamefont{Juge}},
  \bibinfo{author}{\bibfnamefont{S.-G.} \bibnamefont{Je}},
  \bibinfo{author}{\bibfnamefont{D.}~\bibnamefont{de~Souza~Chaves}},
  \bibinfo{author}{\bibfnamefont{S.}~\bibnamefont{Pizzini}},
  \bibinfo{author}{\bibfnamefont{L.}~\bibnamefont{Buda-Prejbeanu}},
  \bibinfo{author}{\bibfnamefont{L.}~\bibnamefont{Aballe}},
  \bibinfo{author}{\bibfnamefont{M.}~\bibnamefont{Foerster}},
  \bibinfo{author}{\bibfnamefont{A.}~\bibnamefont{Locatelli}},
  \bibinfo{author}{\bibfnamefont{T.}~\bibnamefont{Mente{\c{s}}}},
  \bibinfo{author}{\bibfnamefont{A.}~\bibnamefont{Sala}}, \bibnamefont{et~al.},
  \bibinfo{journal}{J. Magn. Magn. Mater.} \textbf{\bibinfo{volume}{455}},
  \bibinfo{pages}{3} (\bibinfo{year}{2018}).

\bibitem[{\citenamefont{Vansteenkiste et~al.}(2014)\citenamefont{Vansteenkiste,
  Leliaert, Dvornik, Helsen, Garcia-Sanchez, and
  Van~Waeyenberge}}]{Vansteenkiste2014}
\bibinfo{author}{\bibfnamefont{A.}~\bibnamefont{Vansteenkiste}},
  \bibinfo{author}{\bibfnamefont{J.}~\bibnamefont{Leliaert}},
  \bibinfo{author}{\bibfnamefont{M.}~\bibnamefont{Dvornik}},
  \bibinfo{author}{\bibfnamefont{M.}~\bibnamefont{Helsen}},
  \bibinfo{author}{\bibfnamefont{F.}~\bibnamefont{Garcia-Sanchez}},
  \bibnamefont{and}
  \bibinfo{author}{\bibfnamefont{B.}~\bibnamefont{Van~Waeyenberge}},
  \bibinfo{journal}{AIP advances} \textbf{\bibinfo{volume}{4}},
  \bibinfo{pages}{107133} (\bibinfo{year}{2014}).

\bibitem[{\citenamefont{Xia et~al.}(2020)\citenamefont{Xia, Zhang, Ezawa, Shao,
  Liu, and Zhou}}]{Xia2020}
\bibinfo{author}{\bibfnamefont{J.}~\bibnamefont{Xia}},
  \bibinfo{author}{\bibfnamefont{X.}~\bibnamefont{Zhang}},
  \bibinfo{author}{\bibfnamefont{M.}~\bibnamefont{Ezawa}},
  \bibinfo{author}{\bibfnamefont{Q.}~\bibnamefont{Shao}},
  \bibinfo{author}{\bibfnamefont{X.}~\bibnamefont{Liu}}, \bibnamefont{and}
  \bibinfo{author}{\bibfnamefont{Y.}~\bibnamefont{Zhou}},
  \bibinfo{journal}{Appl. Phys. Lett.} \textbf{\bibinfo{volume}{116}},
  \bibinfo{pages}{022407} (\bibinfo{year}{2020}).

\bibitem[{\citenamefont{Fruchart et~al.}(2011)\citenamefont{Fruchart, Rousseau,
  Schmaus, L'Hoir, Haettel, and Ortega}}]{Fruchart2011}
\bibinfo{author}{\bibfnamefont{O.}~\bibnamefont{Fruchart}},
  \bibinfo{author}{\bibfnamefont{A.}~\bibnamefont{Rousseau}},
  \bibinfo{author}{\bibfnamefont{D.}~\bibnamefont{Schmaus}},
  \bibinfo{author}{\bibfnamefont{A.}~\bibnamefont{L'Hoir}},
  \bibinfo{author}{\bibfnamefont{R.}~\bibnamefont{Haettel}}, \bibnamefont{and}
  \bibinfo{author}{\bibfnamefont{L.}~\bibnamefont{Ortega}},
  \bibinfo{journal}{Appl. Phys. Lett.} \textbf{\bibinfo{volume}{98}},
  \bibinfo{pages}{131906} (\bibinfo{year}{2011}).

\bibitem[{\citenamefont{Damon and Eshbach}(1961)}]{Damon1961}
\bibinfo{author}{\bibfnamefont{R.}~\bibnamefont{Damon}} \bibnamefont{and}
  \bibinfo{author}{\bibfnamefont{J.}~\bibnamefont{Eshbach}},
  \bibinfo{journal}{J. Phys. Chem. Solids} \textbf{\bibinfo{volume}{19}},
  \bibinfo{pages}{308} (\bibinfo{year}{1961}).

\bibitem[{\citenamefont{Zakeri et~al.}(2010)\citenamefont{Zakeri, Zhang,
  Prokop, Chuang, Sakr, Tang, and Kirschner}}]{Zakeri2010}
\bibinfo{author}{\bibfnamefont{K.}~\bibnamefont{Zakeri}},
  \bibinfo{author}{\bibfnamefont{Y.}~\bibnamefont{Zhang}},
  \bibinfo{author}{\bibfnamefont{J.}~\bibnamefont{Prokop}},
  \bibinfo{author}{\bibfnamefont{T.-H.} \bibnamefont{Chuang}},
  \bibinfo{author}{\bibfnamefont{N.}~\bibnamefont{Sakr}},
  \bibinfo{author}{\bibfnamefont{W.}~\bibnamefont{Tang}}, \bibnamefont{and}
  \bibinfo{author}{\bibfnamefont{J.}~\bibnamefont{Kirschner}},
  \bibinfo{journal}{Phys. Rev. Lett.} \textbf{\bibinfo{volume}{104}},
  \bibinfo{pages}{137203} (\bibinfo{year}{2010}).

\bibitem[{\citenamefont{Di et~al.}(2015)\citenamefont{Di, Zhang, Lim, Ng, Kuok,
  Yu, Yoon, Qiu, and Yang}}]{Di2015}
\bibinfo{author}{\bibfnamefont{K.}~\bibnamefont{Di}},
  \bibinfo{author}{\bibfnamefont{V.~L.} \bibnamefont{Zhang}},
  \bibinfo{author}{\bibfnamefont{H.~S.} \bibnamefont{Lim}},
  \bibinfo{author}{\bibfnamefont{S.~C.} \bibnamefont{Ng}},
  \bibinfo{author}{\bibfnamefont{M.~H.} \bibnamefont{Kuok}},
  \bibinfo{author}{\bibfnamefont{J.}~\bibnamefont{Yu}},
  \bibinfo{author}{\bibfnamefont{J.}~\bibnamefont{Yoon}},
  \bibinfo{author}{\bibfnamefont{X.}~\bibnamefont{Qiu}}, \bibnamefont{and}
  \bibinfo{author}{\bibfnamefont{H.}~\bibnamefont{Yang}},
  \bibinfo{journal}{Phys. Rev. Lett.} \textbf{\bibinfo{volume}{114}},
  \bibinfo{pages}{047201} (\bibinfo{year}{2015}).

\bibitem[{\citenamefont{Nembach et~al.}(2015)\citenamefont{Nembach, Shaw,
  Weiler, Jue, and Silva}}]{Nembach2015}
\bibinfo{author}{\bibfnamefont{H.~T.} \bibnamefont{Nembach}},
  \bibinfo{author}{\bibfnamefont{J.~M.} \bibnamefont{Shaw}},
  \bibinfo{author}{\bibfnamefont{M.}~\bibnamefont{Weiler}},
  \bibinfo{author}{\bibfnamefont{E.}~\bibnamefont{Jue}}, \bibnamefont{and}
  \bibinfo{author}{\bibfnamefont{T.~J.} \bibnamefont{Silva}},
  \bibinfo{journal}{Nature Phys.} \textbf{\bibinfo{volume}{11}},
  \bibinfo{pages}{825} (\bibinfo{year}{2015}).

\bibitem[{\citenamefont{Stashkevich et~al.}(2015)\citenamefont{Stashkevich,
  Belmeguenai, Roussign{\'e}, Cherif, Kostylev, Gabor, Lacour, Tiusan, and
  Hehn}}]{Stashkevich2015}
\bibinfo{author}{\bibfnamefont{A.}~\bibnamefont{Stashkevich}},
  \bibinfo{author}{\bibfnamefont{M.}~\bibnamefont{Belmeguenai}},
  \bibinfo{author}{\bibfnamefont{Y.}~\bibnamefont{Roussign{\'e}}},
  \bibinfo{author}{\bibfnamefont{S.}~\bibnamefont{Cherif}},
  \bibinfo{author}{\bibfnamefont{M.}~\bibnamefont{Kostylev}},
  \bibinfo{author}{\bibfnamefont{M.}~\bibnamefont{Gabor}},
  \bibinfo{author}{\bibfnamefont{D.}~\bibnamefont{Lacour}},
  \bibinfo{author}{\bibfnamefont{C.}~\bibnamefont{Tiusan}}, \bibnamefont{and}
  \bibinfo{author}{\bibfnamefont{M.}~\bibnamefont{Hehn}},
  \bibinfo{journal}{Phys. Rev. B} \textbf{\bibinfo{volume}{91}},
  \bibinfo{pages}{214409} (\bibinfo{year}{2015}).

\end{thebibliography}

\end{document}